# Effects of Low–Pressure Condition on Partial Discharges in WBG Power Electronics Modules

Moein Borghei and Mona Ghassemi
ECE Department, Virginia Polytechnic Institute and State University
Blacksburg, VA, USA

*Abstract*—The aviation industry aims to reduce $CO_2$ emission by reducing energy consumption and benefitting from more electrical systems than those based on fossil fuels. The more electric aircraft (MEA) can take advantage of the drives based on wide bandgap (WBG)-based power modules that are lighter and can bear higher voltages and currents. However, the fast-rise and repetitive voltage pulses generated by WBG-based systems can endanger the insulating property of dielectric materials due to the partial discharges. PDs cause the local breakdown of insulation materials in regions with high electric field magnitude. In the case of power electronic modules, silicone gel, which is a predominant type of encapsulation material, is susceptible to these PDs. In this study, it is aimed to investigate the impact of rise time on various PD characteristics including PD true charge magnitude, inception and extinction fields, duration, and so forth. Besides, those systems that are anticipated to operate under harsh environmental conditions – such as naval or aviation industries – may expect additional threat. Therefore, this paper puts forth the combination of low-pressure conditions and fast-rise, high-frequency square wave pulses. The results demonstrate that more intense and more intense ones are going to be imposed at lower pressures. COMSOL Multiphysics interfaced with MATLAB is used to simulate the PD detection process based on the experimental data found in the literature.

## I. INTRODUCTION

While the aviation industry is expected to meet an increasing demand [1], the number of passengers is going to be doubled by 2040 [2]. This industry must also respond to environmental and economic concerns. This can justify the trend among the big manufacturers in this industry to move forward to more electric aircraft [3].

Although WBG-based systems are the major focus of researchers to achieve all-electric powertrains with higher efficiency and power density, the concern on the reliability of the insulation systems when being exposed to extremely high electric tensions has not been well-responded. In literature, the detrimental impacts of high-frequency, fast-rise voltage pulses generated by WBG-based systems have been reported [4-6].

Besides, the harsh environmental conditions during a flight even make the situation worse for the insulation system. For instance, an aircraft can undergo a pressure as low as 4 psi when flying at cruising altitude where the impact of pressure on PD occurrence has not been sufficiently studied so far. While the impact of different pressure levels (20kPa – 100kPa) on the repetitive partial discharge inception voltage (RPDIV) of a solid dielectric (a twisted wire pair) is investigated in [7], authors in [8], studied the impact of pressure on various characteristics of PD phenomenon by FEA modeling.

This study aims to assess the combined effect of low-pressure levels with ultra-short rise times with the aid FEA model. Therefore, the developed PD model is used to monitor the impact of rise time and low-pressure level on the PD intensity and duration, simultaneously.

## II. PD MECHANISM

### A. PD Inception

The electric field threshold for a local breakdown (PD) is different from that of the global breakdown. However, the PD inception field is directly related to the critical electric field magnitude for global breakdown [9]. It is also a function of the properties of the gas inside the bubble as well as the bubble size. The following equation yields the PD inception electric field which is derived from the streamer inception criterion [9]:

$$E_{inc} = (E/p)_{cr}\, p \left(1 + \frac{B}{(pl)^{0.5}}\right) \qquad (1)$$

In the above expression, $(E/p)_{cr}$, $B$, and $n$ are gas parameters. In the air, their values are $24.2\ V\ Pa^{-1}m^{-1}$, $8.6\ Pa^{0.5}m^{0.5}$, and $0.5$, respectively. The pressure of air is denoted as $p$, while $l$ represents the cavity diameter.

Aside from the above condition, there is another condition that must be met to have an ionization taking place. It is the existence of initial free electron(s); this condition stands for the statistical time lag before the occurrence of a PD and is stochastic in nature. There are various mechanisms for the provision initial electron that are explained in [10]; the major source for initial electrons are known to be the detrapping of electrons produced during previous PD activities and accumulated on the cavity surface. In this paper, to truly examine the impact of rise time and pressure, the stochastic intervention of electron generation processes is neglected.

### B. PD Extinction

When a PD activity ignites, the collisions among electrons and atoms keep happening until a moment where the electron does not have enough energy to further expand the streamer. Same as the PD inception criterion, the PD extinction criterion is also a field-dependent condition. The electric field at which streamer cannot further grow and PD quenches is called the *Extinction Electric Field* ($E_{ext}$). According to Gutfleisch and Niemeyer [11, 12], $E_{ext}$ can be obtained as:

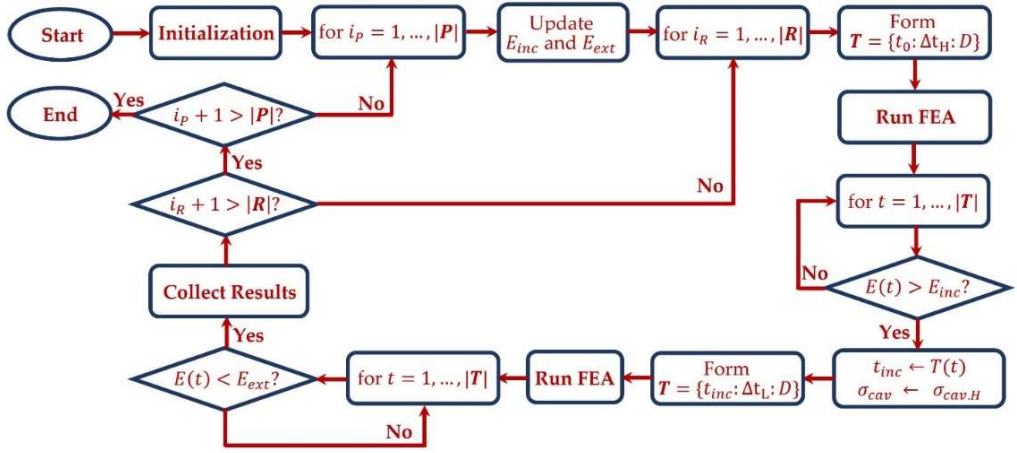

Fig. 1. The flowchat for assessing the impact of air pressure and rise time on PD characteristics

$$E_{ext} = \gamma\, p\, (E/p)_{cr} \qquad (2)$$

In the above expression, $\gamma$ is a dimensionless factor varying with the change of voltage polarity. $\gamma$ can be estimated experimentally. This work determines $\gamma$ such that at normal temperature and pressure (NTP) conditions ($p_0, T_0$), $E_{ext}$ equals the value proposed in the literature.

*C. PD Transient State*

During the PD activity, the reason behind the decrease in the electric field inside the air-bubble is the higher conductivity of the cavity compared to its normal condition. Starting from the inception time of a PD, the conductivity of cavity increases from an initial value (approximately zero) to a higher value ($\sigma_{cav,max}$).

Between the time at which a PD starts ($t_{inc}$), and the time at which PD quenches ($t_{ext}$), the electric charges (electrons and positive ions) are formed and accumulate over the cavity and electrode surfaces. The amount of charge accumulated on each upper/lower surface of the cavity is called *PD true charge magnitude* ($q_{true}$). The value of $q_{true}$ is a parameter for demonstrating the intensity of a PD event:

$$q_{true} = \int_{t_{inc}}^{t_{ext}} I_{cav}(t)\, dt \qquad (3)$$

where

$$I_{cav}(t) = \int_{S_{cav}} J(t)\, dS \qquad (4)$$

In the above expressions, $J(t)$ is the magnitude of current density, and $I_{cav}(t)$ is the magnitude of current flow through the upper/lower surface of the void ($S_{cav}$).

### III. ROLE OF PRESSURE

*A. Dielectric Constant (relative permittivity)*

To integrate the impact of low-pressure conditions with high electric tension due to the applied voltage, this paper investigates the role of pressure in three parameters:

(1) PD inception field
(2) PD extinction field
(3) Dielectric constant

Among the above three parameters, the dependencies of $E_{inc}$ and $E_{ext}$ on pressure are shown in (1) and (2). For dielectric constant– which plays a prominent role in the propagation of electric field–, authors examined the dependency of $\epsilon_r$ on pressure for silicone gel [8]. This relationship can be expressed by a Tait-like expression [13]:

$$1 - \frac{\epsilon_{p_0}}{\epsilon_p} = A \ln\left(\frac{B+p}{B+p_0}\right) \qquad (3)$$

where $\epsilon_{p_0}$ and $p_0$ denote the dielectric constant and pressure at sea level and room temperature. Similarly, $\epsilon_p$ is the dielectric constant at pressure $p$. The constants $A$ and $B$ are functions of the material and must be experimentally determined. Approximate values for $A$ and $B$ are derived from [14] in this work.

Generally, the dielectric constant has a direct relationship with pressure meaning that as the pressure declines, $\epsilon_r$ also goes down. However, according to (5), the rate of change declines as pressure is reduced and $\epsilon_r$ has little variations for $p < 1$ atm.

### IV. PD MODELING

In this study, COMSOL Multiphysics and MATLAB are used for modeling the PD phenomenon as well as performing the sensitivity analysis for parameters of interest. COMSOL Multiphysics takes over the part of the finite-element analysis which is necessary for electric field estimation while MATLAB is used to model the algorithm of PD detection.

Figure 1 shows the flowchart for PD modeling which also demonstrates the procedure of conducting sensitivity analysis for pressure and rise time. It starts with initialization step where the parameters associated with the model are valued. Then, at each pressure level to be evaluated, a list of rise times is considered. The values of $E_{inc}$ and $E_{ext}$ at each pressure level are updated. Then, for each rise-time value, the time duration under consideration ($D$) is discretized by time steps $\Delta t_H$ and

$\Delta t_L$ which are attributed to normal conditions (no-discharge period) and during PD activity. These time steps vary depending on the rise time value; the shorter the rise time is, the smaller the time steps will be. Prior to the PD occurrence, the time set, **T**, is formed based on $\Delta t_H$, and then, the FEA is run. Starting from the first-time step, the time steps are evaluated by streamer inception criterion to see if PD happens. The first-time step that satisfies the aforementioned condition is the PD inception time ($t_{inc}$) and the conductivity of the cavity is updated to $\sigma_{cav,H}$. Then, the FEA is run again, this time with a time step that starts from $t_{inc}$ and goes on by $\Delta t_L$. Then, the first-time step that satisfies the PD extinction condition indicates the PD extinction time ($t_{ext}$). The induced charge on the cavity surface between $t_{inc}$ and $t_{ext}$ is the PD true charge. This process is repeated until all the rise times for all the pressure levels are investigated.

## V. Results

The geometry that was examined for the modeling and simulations is shown in Fig. 2 and consists of two spherical electrodes; one is the high-voltage electrode while the other is grounded. The two spheres are situated within a cylindrical block of silicone gel ($\varepsilon_r = 2.7$ at normal temperature and pressure conditions) with a spherical air-filled cavity inside. The geometric parameters of the void-dielectric set are shown in Table I.

Due to the symmetry of the configuration, a 2D axisymmetrical representation of geometry is used for electric field estimation.

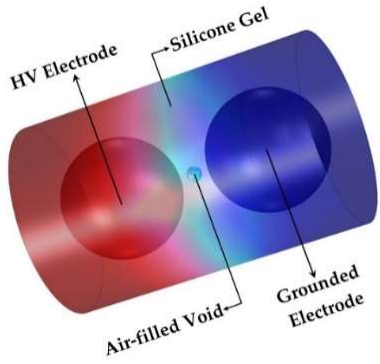

Fig. 2. The geometry of the case study.

TABLE I. DIMENSIONS OF THE CASE STUDY

| Parameter | Dielectric Block Height | Dielectric Radius | Cavity Diameter | Electrode Diameter |
|---|---|---|---|---|
| Value | 25 mm | 8 mm | 1.2 mm | 10 mm |

The properties of the unipolar square wave voltage for the base case are shown in Table II.

TABLE II. UNIPOLAR SQUARE WAVE VOLTAGE PARAMETERS

| Parameter | $U_{max}$ | f | Rise time | Duty Cycle |
|---|---|---|---|---|
| Value | 18 kV | 10 kHz | 50 ns | 50% |

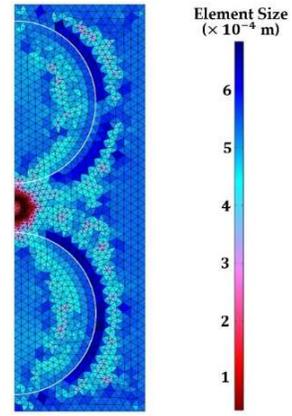

Fig. 3. The mesh pattern of the case study (2D axisymmetrical).

Figure 3 shows the mesh pattern for the 2D axisymmetrical representation of the case study. The electrode boundaries are highlighted by white lines and as can be seen, the density of very small triangular elements is much higher in the periphery of the cavity. The reason is the higher field intensity in this region which demands a higher resolution of FEA study. The numerical properties of mesh patterns are brought in Table III.

TABLE III. MESH PROPERTIES

| Parameter | Maximum element size | Minimum element size | Maximum element growth rate | Curvature factor |
|---|---|---|---|---|
| Value | $5.3 \times 10^{-4}$ m | $3 \times 10^{-6}$ m | 1.3 | 0.3 |

Next, the results of examining the air pressure and rise impacts on PD characteristics are reported. In this regard, several rise time values in the range of $1\ ns - 1\ \mu s$ are considered where the focus is mostly on ultra-short rise times of about nanoseconds. The sensitivity analysis of rise time is performed at five different pressure levels from $0.2\ atm$ to $1\ atm$ which helps us investigate the low-pressure condition impact.

Fig. 4 shows the variations of $E_{inc}$ and $E_{ext}$ with pressure. As seen, these two parameters have opposite relationships with pressure change. While $E_{inc}$ goes higher at low pressures, $E_{ext}$ is an increasing function of pressure. As we will see the impact of these variations in the PD results, these trends cause the prolongation of PD events while making them more intense.

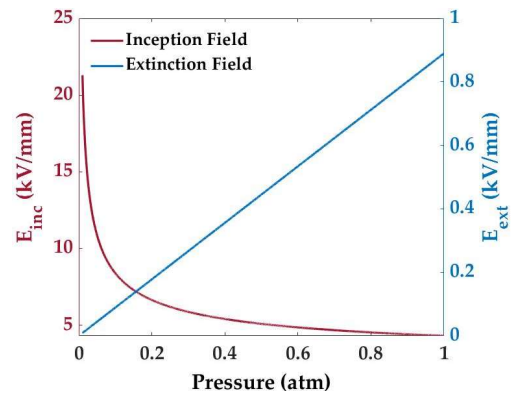

Fig. 4. The PD inception and extinction of electric fields versus pressure.

Figure 5 demonstrates the variations of PD duration as a function rise time at different pressures. The results imply that there is not a strict pattern for the changes, but a longer PD event is expected at shorter rise times. Moreover, the pressure reduction harms PD duration meaning that as the pressure declines, the PD events tend to be longer. This observation can be justified by the trend seen in Figure 4; a higher $E_{inc}$ mandates the PD to take more time for reaching the inception field while $E_{ext}$ do similar for the extinction time.

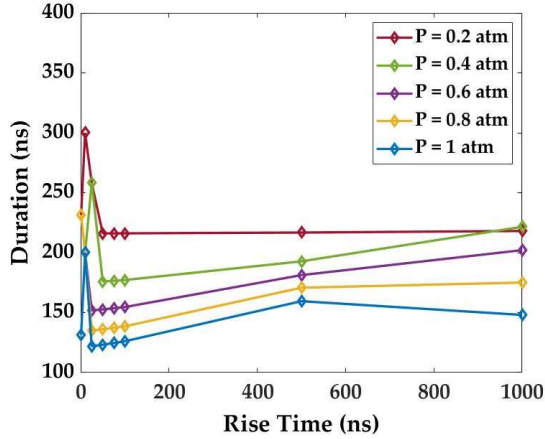

Fig. 5. The relationship between PD duration and rise time at different pressures.

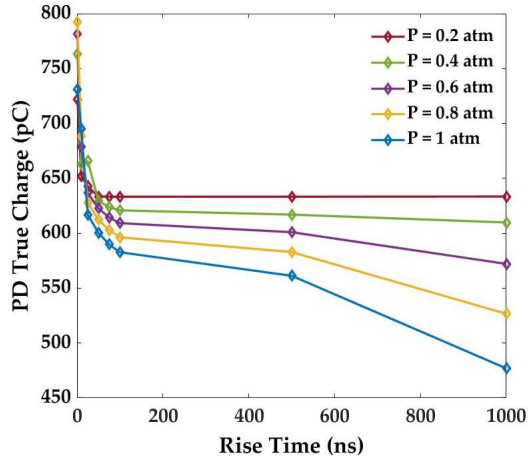

Fig. 6. Variations of PD true charge as a function of pressure.

Fig. 6 shows variations of PD true charge magnitude versus rise time at different pressure levels. As expected, lower rise time imposes a more intense discharge activity. The things exacerbate when operating at lower pressure levels. There are three reasons for that: (1) discharge occurs at higher field magnitudes which in turn, causes the current density to be higher (Ohm's law), (2) according to $D = \epsilon E$, the field displacement goes down at lower dielectric constants. As discussed earlier in section III, at lower pressures, the dielectric constant declines according to the Taie-like expression. As the field displacement has directly affected the true charge magnitude, at a lower dielectric constant, the amount of true charge would be higher.

## VI. CONCLUSION

The electrification trend in the aviation industry requires the reliability of the electrified system in all aspects. Therefore, the threats imposed on the insulation system by the WBG-based power modules are studied in this work. Because the pressure at higher altitudes can be extremely low, this study examined the combined impact of high slew rate and low-pressure conditions with the aid of the FEA model. The results show that while short rise time square wave voltages are expected to cause more intense discharges with longer durations, the low-pressure conditions make these harmful changes even worse. The main reasons behind these changes are the changes in the inception and extinction electric field magnitudes as well as the lower dielectric constant values at higher altitudes.